\documentclass[1p]{elsarticle}
\usepackage{mathtools,amsfonts,amssymb,amsthm}
\usepackage{times,graphicx,xcolor}
\usepackage{delarray,fancybox}
\usepackage{mathdots}

\newcommand{\dn}{d_{n}}

\newcommand{\tr}{\mathrm{tr}}
\newcommand{\qc}[2]{\mathcal{C}_{S#1(d)}^{(2,#2)}}
\newcommand{\qcu}[2]{\mathcal{C}_{SU(#1)}^{(2,#2)}}
\newcommand{\ud}[1]{\mathfrak{g}_{#1}}

\newcommand{\ivec}{\mathcal{T}}
\title{Towards a geometrical classification of statistical conservation laws 
in turbulent advection.}
\date{}
\author[els]{Paolo Muratore-Ginanneschi}
\ead[url]{http:http://mathstat.helsinki.fi/mathphys/paolo.html}
\address[els]{Department of Mathematics and Statistics, 
University of Helsinki, PB 68 Helsinki 00014, Finland}
\begin{document}
\begin{abstract}
The paper revisits the compressible Kraichnan model of turbulent advection
in order to derive explicit quantitative relations between scaling 
exponents and Lagrangian particle configuration geometry.
\end{abstract}
\maketitle

\section{Introduction}
\label{sec:intro}

Physical and numerical experiments have typically access to, at least,
two kinds of observables of the state of a turbulent Newtonian fluid.
These observables are, on the one hand, Eulerian or Lagrangian statistical
indicators such as universal scaling properties of correlation and structure
functions. On the other hand, geometrical indicators like shapes formed
by configurations privileged by the dynamics of Lagrangian or inertial particles
can be observed. A series of remarkable results (see e.g. \cite{FaSr06,Fa09} 
and references therein) have in recent years started to unveil the nature of     
the link between statistical indicators and geometrical properties. 
In particular, \cite{CeVe01} gave clear evidence that tracer particles advected
by the $2d$ Navier--Stokes velocity field spend anomalously long times in degenerate geometries 
characterized by strong clustering and that the phenomenon has a quantitative
counterpart in the existence of special functions of particle configurations which are 
on average preserved by flow. Furthermore, it was shown in \cite{BeGaHo04}
for densities transported by compressible turbulent velocity fields that 
statistical conservation laws determine, at least for integer values of 
the mass, the multifractal spectrum of the attractor towards which 
Lagrangian trajectories converge. 
On the background of these results was the discovery of anomalous scaling
\cite{GaKu95,ChFaKoLe95,ShSi95} in a stylized model of passive advection, 
the Kraichnan model \cite{Kr68}. There it was possible to show in a mathematically 
controlled fashion \cite{BeGaKu98}, that the universal statistical properties 
of the model in the inertial and, under additional hypotheses \cite{FaFo05}, 
decay ranges are determined by statistical conservation laws of the 
Lagrangian dynamics (referred to as zero modes). 
Zero modes have subsequently become the object of extensive investigations and 
analytical, mostly perturbative, and numerical expressions of their scaling
exponents have been derived for the Kraichnan model and its generalizations 
(see e.g. \cite{FaGaVe01,KuMG07} and references therein). 
To the best of my knowledge, however, quantitative expressions 
of the scaling exponents explicitly relating them to geometrical properties
of Lagrangian particle configurations appeared only recently in \cite{MaMG09}.
The scope of the present contribution is to illustrate the new method of calculation
introduced in \cite{MaMG09} in the slightly more general case of the compressible 
Kraichnan model \cite{GaVe00}. The main result of the paper, derived in sections~\ref{sec:mg} and 
\ref{sec:csd}, is the classification of zero modes, irreducible 
and reducible, in terms of Gel'fand-Zetlin patterns (see e.g. \cite{Lo70}) 
associated to quadratic Casimir operators of classical groups. It must be clearly stated 
that this classification is perturbative but it has nevertheless the merit 
to establish an explicit link to Lagrangian particle geometry which is discussed 
in more details for four point correlations in section~\ref{sec:duality}. 
Sections~\ref{sec:model} and \ref{sec:Hopf} recall respectively the defining 
properties of the Kraichnan model and basic facts about the structure of 
the solution of the Hopf equations.
In the conclusions I discuss the use of the new method and make some speculations 
on its significance for the Navier--Stokes equation as well as for other 
statistical field theories.

\section{The compressible Kraichnan model}
\label{sec:model}

Passive turbulent advection by a \emph{compressible} velocity field 
in $\mathbb{R}^{d}$ encompasses (see e.g. \cite{GaVe00}) the density evolution 
equation
\begin{eqnarray}
\label{model:density}
\partial_{t}\rho+\partial_{\boldsymbol{x}}\cdot(\boldsymbol{v} \rho)=
\frac{\kappa}{2} \partial_{\boldsymbol{x}}^{2}\rho +f
\end{eqnarray}
(describing e.g. the distribution of particles floating on the surface of the
fluid in the limit of vanishing Stokes number) and the tracer evolution equation
\begin{eqnarray}
\label{model:tracer}
\partial_{t}\theta+\boldsymbol{v}\cdot\partial_{\boldsymbol{x}}\theta=
\frac{\kappa}{2} \partial^{2}\theta +g
\end{eqnarray}
(describing e.g. the fluid temperature). At finite molecular viscosity
$\kappa$, the Lagrangian dynamics underlying (\ref{model:density}) and 
(\ref{model:tracer}) are distinct, respectively forward and backward in time. 
In the \emph{inviscid limit} and in the case
of advection by a turbulent field stylized by a stationary, time-decorrelated
Gaussian ensemble they become formally adjoint \cite{GaVe00} and can be 
discussed in parallel. These are the working hypotheses of the paper.
In particular, the velocity field in (\ref{model:density}), (\ref{model:tracer})
is modeled by the realizations of Kraichnan's zero average Gaussian 
compressible ensemble: 
\begin{subequations}
\begin{eqnarray}
\label{}
\prec  v^{\alpha}(\boldsymbol{x}_{1},t_{1})v^{\beta}(\boldsymbol{x}_{2},t_{2})\succ 
=\delta(t_{12})D^{\alpha\,\beta}(\boldsymbol{x}_{12},m)
\end{eqnarray}
\begin{eqnarray}
\label{}
D^{\alpha\,\beta}(\boldsymbol{x};m)=D_{0}\,\xi\,(2-\xi)\int \frac{d^{d}p}{(2\,\pi)^{d}}
\frac{e^{\imath \boldsymbol{p}\cdot \boldsymbol{x}}}{p^{d+\xi}} 
\left\{ (1-\wp)\delta^{\alpha\,\beta}-(1-d\,\wp) 
\frac{p^{\alpha}p^{\beta}}{p^2}\right\}\chi\left(\frac{m^{2}}{p^{2}}\right)
\end{eqnarray}
\end{subequations}
where $\boldsymbol{x}_{12}:=\boldsymbol{x}_{1}-\boldsymbol{x}_{2}$, 
$t_{12}:=t_{1}-t_{2}$, $\xi\in[0,2]$ is the roughness degree of the 
velocity field,
$\wp\in[0,1]$ is the degree of compressibility and $\chi$ some 
non-universal infra-red cut-off function normalized to the unity for vanishing
inverse integral scale $m$. The force fields $f$, $g$ 
(Gaussian, zero-average and decorrelated in time) compensate the dissipation,
in the inviscid limit only due to the eddy diffusivity generated by the
velocity field, in order to let the system attain a Galilean invariant 
steady state.
Consistence with hydrodynamics imposes to interpret
(\ref{model:density}), (\ref{model:tracer}) advected by the Kraichnan
ensemble as stochastic partial differential equations in the sense of 
Stratonovich. The role of the eddy diffusivity appears upon converting 
(\ref{model:density}), (\ref{model:tracer}) into Ito form 
\cite{FaGaVe01,KuMG07}. This latter is for the density
\begin{eqnarray}
\partial_{t}\rho+\partial_{\boldsymbol{x}}\cdot(\boldsymbol{v}   \rho)
=\frac{\varkappa\,m^{-\xi}}{2} \partial^{2}\rho +f
\end{eqnarray}
and for the tracer
\begin{eqnarray}
\partial_{t}\theta+\boldsymbol{v}\cdot\partial_{\boldsymbol{x}}\theta
=\frac{\varkappa\,m^{-\xi}}{2} \partial^{2}\theta +g
\end{eqnarray}
The eddy diffusivity in the above equations is specified by
\begin{eqnarray}
\varkappa\,m^{-\xi}=\frac{D^{\alpha}_{\hspace{0.1cm}\alpha}(\boldsymbol{0},m)}{d}
=\frac{d-1}{d}\int\frac{d^{d}p}{(2\pi)^{d}}\frac{D_{0}\,\xi}{p^{d+\xi}}
\chi\left(\frac{m^{2}}{p^{2}}\right) 
\end{eqnarray}
which means that Taylor's formula \cite{Ta22} becomes exact in the 
Kraichnan model.

\section{Hopf equations and martingales}
\label{sec:Hopf}

A straightforward application of Ito calculus to (\ref{model:tracer})
shows that the equal time correlation $\mathcal{C}_{n}$ of $n$ tracer fields 
in $\mathbb{R}^{d}$ satisfies the Hopf equation (see e.g. \cite{BeGaKu98, FaGaVe01})
\begin{eqnarray}
\label{Hopf:Hopf}
\left(\partial_{t}+\mathcal{M}_{\boldsymbol{X}}^{(n)}\right) 
\mathcal{C}_{n}\left(\boldsymbol{X};t\right)
=\mathcal{F}_{n}\left(\boldsymbol{X};t\right)
\end{eqnarray}
with
\begin{eqnarray}
\label{Hopf:M_n}
\mathcal{M}_{\boldsymbol{X}}^{(n)}=-
\frac{1}{2}\sum_{ij}^{n}
D^{\alpha\,\beta}(\boldsymbol{x}_{ij},m)
\partial_{x_{i}^{\alpha}}\partial_{x_{j}^{\beta}}
\end{eqnarray}
and $\mathcal{F}_{n}$ an effective forcing term fully 
specified by the correlation of $g$ and by correlation functions 
$\mathcal{C}_{n^{\prime}}$ with $n^{\prime}\,<\,n$. The operator $\mathcal{M}_{n}$ can be also regarded as the generator 
of a multiplicative diffusion process the transition probability $\mathcal{P}_{n}$ 
whereof obeys 
\begin{subequations}
\begin{eqnarray}
\label{}
\label{Hopf:tracer}
\left(\partial_{t}+\mathcal{M}_{\boldsymbol{X}}^{(n)}\right) 
\mathcal{P}_{n}\left(\boldsymbol{X},\boldsymbol{X}^{\prime};t-t^{\prime}\right)
=0 
\end{eqnarray}
\begin{eqnarray}
\label{}
\lim_{t\downarrow t^{\prime}}
\mathcal{P}_{n}\left(\boldsymbol{X},\boldsymbol{X}^{\prime};t-t^{\prime}\right)=
\delta^{(n\,d)}(\boldsymbol{X}-\boldsymbol{X}^{\prime})
\end{eqnarray}
\end{subequations}

The Lagrangian interpretation of $\mathcal{P}_{n}$ is that it is the probability 
density to find a cluster of $n$ inertial particles in 
 $\boldsymbol{X}^{\prime}=(\boldsymbol{x}_{1}^{\prime},
\dots,\boldsymbol{x}_{n}^{\prime})$ at time $t^{\prime}$ conditioned upon the event that they reach
$\boldsymbol{X}=(\boldsymbol{x}_{1},\dots,\boldsymbol{x}_{n})$
at a later time $t$. 
Analogously, the equal time $n$-point correlation of the density field
satisfies an Hopf equation formally adjoint to (\ref{Hopf:Hopf}). 
In such a case $\mathcal{P}_{n}$ can be regarded as the solution of
the Fokker-Planck equation
\begin{eqnarray}
\label{Hopf:density}
\left(\partial_{t}+\mathcal{M}_{\boldsymbol{X}^{\prime}}^{(n)\dagger}\right) 
\mathcal{P}_{n}\left(\boldsymbol{X},\boldsymbol{X}^{\prime};t-t^{\prime}\right)
=0
\end{eqnarray}
with
\begin{eqnarray}
\label{}
\mathcal{M}_{\boldsymbol{X}^{\prime}}^{(n)\dagger}=-
\frac{1}{2}\sum_{ij}^{n}\partial_{x_{i}^{\alpha}}\partial_{x_{j}^{\beta}}
D^{\alpha\,\beta}(\boldsymbol{x}_{ij},m)
\end{eqnarray}
 $\mathcal{P}_{n}$ acquires the interpretation of the probability density 
for a cluster of $n$ inertial particle to \emph{arrive} in
 $\boldsymbol{X}^{\prime}=(\boldsymbol{x}_{1}^{\prime},
\dots,\boldsymbol{x}_{n}^{\prime})$ at time $t$ conditioned upon the event that they are in
 $\boldsymbol{X}=(\boldsymbol{x}_{1},\dots,\boldsymbol{x}_{n})$ at time $t^{\prime}\,\leq\,t$.
The different interpretations of (\ref{Hopf:tracer}) and (\ref{Hopf:density}) stem 
from the distinct Lagrangian dynamics underlying (\ref{model:density}), (\ref{model:tracer}) 
at finite molecular viscosity \cite{GaVe00}. Whenever the kernel $\mathcal{P}_{n}$ acts on 
translational invariant functions, the arithmetic average of the Lagrangian 
particle positions is integrated out. The reduction defines a $d_{n}=(n-1)\,d$-dimensional
subspace of $\mathbb{R}^{n\,d}$, the translational invariant sector of the theory. 
The projection $P_{n}$ of $\mathcal{P}_{n}$ on the subspace has a well defined scale invariant limit 
for $m$ tending to zero \cite{BeGaKu98}. 
The scaling dimensions of $P_{n}$ corresponding to a linear rescaling 
of spatial variables (i.e. in units $\mathsf{d}_{x}$) are 
\begin{eqnarray}
\label{}
\mathsf{d}_{t}=(2-\xi)\,\mathsf{d}_{x}
\hspace{1.0cm}\&\hspace{1.0cm}
\mathsf{d}_{P_{n}}=-(n-1)\,d\,\mathsf{d}_{x}\equiv\,-\,d_{n}\,\mathsf{d}_{x}
\end{eqnarray}
Qualitative analysis \cite{BeGaKu98} yields for $P_{n}$ the asymptotic
expansion:
\begin{eqnarray}
\label{Hopf:expansion}
P_{n}(\boldsymbol{Y},\boldsymbol{Y}^{\prime},t-t^{\prime})
= \left\{\begin{array}{ll}
\sum_{i=0}^{\infty}\phi_{i;1}\left(\boldsymbol{Y}\right)
\psi_{i;1}\left(\boldsymbol{Y}^{\prime},t-t^{\prime}\right) & \mbox{for}\, Y\leq Y^{\prime}
\\[0.2cm]
\sum_{i=0}^{\infty}\psi_{i;2}\left(\boldsymbol{Y},t-t^{\prime}\right)
\phi_{i;2}\left(\boldsymbol{Y}^{\prime}\right)& \mbox{for}\, Y > Y^{\prime}
\end{array}
\right.
\,,\hspace{0.5cm}
\boldsymbol{Y},\boldsymbol{Y}^{\prime}\in\mathbb{R}^{d_{n}}
\end{eqnarray}
Throughout the manuscript, $H_{a}$ denotes the Heaviside function with 
normalization $H_{a}(0)=a$.
The meaning of the expansion is well illustrated by the $\xi=0$ limit. 
In such a case 
\begin{eqnarray}
\mathcal{M}_{\boldsymbol{X}}^{(n)}=\mathcal{M}_{\boldsymbol{X}}^{(n)\dagger}=
-\frac{\varkappa}{2}\partial_{\boldsymbol{X}}^{2}\equiv -\frac{\varkappa}{2}\Delta_{n\,\boldsymbol{X}}
\nonumber
\end{eqnarray}
and both $\mathcal{P}_{n}$ and the reduced propagator $P_{n}$ are Gaussian.
Self-adjointness of the Laplacian also implies
\begin{eqnarray}
\label{}
\psi_{i}=\psi_{i;1}=\psi_{i;2}\hspace{1.0cm}\& \hspace{1.0cm} \phi_{i}=\phi_{i;1}=\phi_{i;2}
\end{eqnarray} 
It is expedient to couch the sum over $i$ into a sum over 
$(\jmath,\boldsymbol{l},k)$, $\jmath$ being the degree of homogeneity
of an harmonic polynomial $\mathcal{H}_{\jmath\boldsymbol{l}}$, $\boldsymbol{l}$ 
the numbers specifying an $SO(d)$-adapted representation 
$\mathcal{Y}_{\jmath\boldsymbol{l}}$ of hyperspherical harmonics of $SO(d_{n})$ 
\cite{FaGrBoHe99} and
\begin{eqnarray}
\label{Hopf:gauss_martingales}
\phi_{\jmath\boldsymbol{l}k}(\boldsymbol{Y}):=
\frac{Y^{2\,k}\,\mathcal{H}_{\jmath\boldsymbol{l}}(\boldsymbol{Y})}
{2^{2k}\,\Gamma(k+1)\,\Gamma\left(\frac{\dn+2\,\jmath+2k}{2}\right)}
=\frac{Y^{2\,k+\jmath}\,\mathcal{Y}_{\jmath\boldsymbol{l}}(\boldsymbol{Y}/Y)}
{2^{2k}\,\Gamma(k+1)\,\Gamma\left(\frac{\dn+2\,\jmath+2k}{2}\right)}
\end{eqnarray}
for $Y\equiv\left|\left|\boldsymbol{Y}\right|\right|$. The so defined  
$\phi_{\jmath\boldsymbol{l}k;1}$'s satisfy the ``tower'' relations
\begin{eqnarray}
\label{Hopf:tower1}
\partial_{\boldsymbol{Y}}^{2}\phi_{\jmath\boldsymbol{l}k}(\boldsymbol{Y})
=\phi_{\jmath\boldsymbol{l}k-1}(\boldsymbol{Y}), \hspace{1.5cm}
\phi_{\jmath\boldsymbol{l}\,-1}(\boldsymbol{Y})=0
\end{eqnarray}
Under the same conventions, the identity 
\begin{eqnarray}
\label{Hopf:ls}
\psi_{\jmath\boldsymbol{l}k}(\boldsymbol{Y},t)
=2\,\pi^{\frac{\dn}{2}}
\frac{H_{\jmath\boldsymbol{l}}(\boldsymbol{Y})\,e^{-\frac{R^{2}}{2\,\varkappa\,t}}\,
L_{k}^{( \frac{\dn+2\,\jmath-2}{2})}\left(\frac{R^{2}}{2\,\varkappa\,t}\right)
\Gamma\left(k+1\right)}{2^{\jmath-k}\,
(2\,\pi\,\varkappa\,t)^{\frac{\dn}{2}}\,(\varkappa\,t)^{\jmath+k}}
\end{eqnarray}
holds true for  $L^{(a)}_{b}$ the generalized Laguerre polynomial 
of degree $k$. An alternative useful representation is
\begin{eqnarray}
\label{Hopf:ho}
\psi_{\jmath\boldsymbol{l}k}(\boldsymbol{Y},t)
=2\,\pi^{\frac{\dn}{2}}
\frac{\left\langle\boldsymbol{Y},\varkappa\,t\right|\left|\jmath\,\boldsymbol{l}\,k\right\rangle\,
e^{-\frac{Y^{2}}{4\,\varkappa\,t}}}{2^{\jmath-k-1}\,(2\,\pi\,\varkappa\,t)^{\frac{\dn}{2}}
\,(\varkappa\,t)^{\jmath+k}}
\end{eqnarray}
with $\left\langle\boldsymbol{Y},\varkappa\,t\right|\left|\jmath\,\boldsymbol{l}\,k\right\rangle$ 
the wave function of the $(\jmath,\boldsymbol{l},k)$ eigenstate 
of the isotropic quantum harmonic oscillator of 
unit mass and frequency $\varkappa\,t$ in $\mathbb{R}^{d_{n}}$\cite{Ro05}. 
Finally, the $\psi_{\jmath\boldsymbol{l}k}(\boldsymbol{Y},t)$'s satisfy the ``tower'' relations
\begin{eqnarray}
\label{Hopf:tower2}
\partial_{t}\psi_{\jmath\boldsymbol{l}k}(\boldsymbol{Y},t)=
\frac{\varkappa}{2}\partial_{\boldsymbol{Y}}^{2}\psi_{\jmath\boldsymbol{l}k}(\boldsymbol{Y},t)=
-\psi_{\jmath\boldsymbol{l}k+1}(\boldsymbol{Y},t)
\end{eqnarray}
Some remarks are in order.
\begin{itemize}
\item[\emph{i}] The identities (\ref{Hopf:tower1}), 
(\ref{Hopf:tower2}) are particular case of consistence conditions 
of the expansion (\ref{Hopf:expansion}) holding for generic $\xi$ \cite{BeGaKu98, FaGaVe01}.
\item[\emph{ii}] By (\ref{Hopf:ho}) and the orthonormality of quantum
eigenstates it readily follows that the harmonic polynomial $\phi_{j\boldsymbol{l}0}$
is $\mathbb{L}^{2}(\mathbb{R}^{d_{n}})$-orthogonal to the $\psi_{\jmath\boldsymbol{l}k}$'s 
for any $k>1$. The normalization of the (\ref{Hopf:gauss_martingales})'s
is chosen such that they are exactly replicated whenever averaged over the
transition probability $P_{n}$. In other words, they are martingales of 
the Wiener process \cite{Oksendal}.
\item[\emph{iii}] The time integral 
\begin{eqnarray}
\label{Hopf:local}
\int_{0}^{\infty}dt\,\psi_{\jmath\,\boldsymbol{l}\,k}^{(0)}(\boldsymbol{Y},t)\propto 
\frac{H_{\jmath\,\boldsymbol{l}}(\boldsymbol{Y})}{Y^{d_{n}+2\,(\jmath+k-1)}}
\end{eqnarray} 
defines for $k$ equal zero only \emph{local}  martingales (i.e. harmonic functions). 
The singularity for $Y\downarrow 0$, obstructs (\ref{Hopf:local}) 
absolute integrability and therefore the martingale property.
This fact has important consequences for perturbative investigation of anomalous scaling 
in the decay range (scales larger than the integral scale 
of the forcing for infinite integral scale of the velocity field \cite{MaMG09}).
\end{itemize}

\section{Scaling, martingales of the diffusion process and perturbation theory}
\label{sec:mg}

The term martingales \cite{Oksendal} denotes functions of 
a Markov process the shape of which, on average, is 
preserved by the time evolution:
\begin{eqnarray}
\label{mg:def}
\int d^{d}x\,\mathcal{F}(\boldsymbol{x},s)\,
K\left(\boldsymbol{x},s|\boldsymbol{x}^{\prime},s^{\prime}\right)
=\mathcal{F}(\boldsymbol{x}^{\prime},s^{\prime})
\end{eqnarray}
The kernel $K$ in (\ref{mg:def}) is a probability density for $\boldsymbol{x}$
at time $s$ conditioned upon the occurrence of $\boldsymbol{x}^{\prime}$ at time
$s^{\prime}$. For forward (backward) dynamics $s>s^{\prime}$
($s^{\prime}>s$). Clearly, a time independent 
$\mathcal{F}(\boldsymbol{x})$ satisfying (\ref{mg:def}) is on average a conserved quantity
of the diffusion process described by the transition probability $K$.
Furthermore, (\ref{mg:def}) requires only integrability of 
$\mathcal{F}(\boldsymbol{x})$ with respect to the transition probability. The functional space where 
statistical conservation laws can be sought is therefore larger than 
that of the physical solutions of the Hopf equations. Turning the attention
to the dual action of $K$, the role of constant martingales is played 
by stationary pseudo-measure $\mathcal{G}$ satisfying
\begin{eqnarray}
\label{mg:pseudo}
\int d^{d}x^{\prime}\,\,
K\left(\boldsymbol{x},s|\boldsymbol{x}^{\prime},s^{\prime}\right)
\mathcal{G}(\boldsymbol{x}^{\prime})=\mathcal{G}(\boldsymbol{x})
\end{eqnarray}
The relevance of these considerations for the Kraichnan model 
\cite{BeGaKu98,FaGaVe01} is that the expansion (\ref{Hopf:expansion}) 
generically comprises constant martingales
$\mathcal{Z}_{i;1}$and stationary pseudo-measures $\mathcal{Z}_{i;2}$. They satisfy the zero mode
equations 
\begin{eqnarray}
\label{mg:zm}
\mathcal{M}^{(n)}\mathcal{Z}_{i;1}=0
\hspace{1.0cm}\&\hspace{1.0cm}
\mathcal{M}^{(n)\dagger}\mathcal{Z}_{i;2}=0
\end{eqnarray}
and asymptotically dominate in the inertial range the scaling
of respectively the tracer and density correlation functions.
Observation $\emph{iii}$ at the end of the previous section evinces
that the concept of statistical conservation law (i.e. martingale or
stationary pseudo-measure) is stronger than the one of zero mode 
as the (\ref{mg:zm})'s provide only necessary conditions.
For small $\xi$, small scale zero modes are deformations of harmonic 
polynomials which are statistical conservation laws of the Wiener process.  
In such a case the identification of zero modes with statistical conservation
laws holds true. The fact suggests to use (\ref{mg:def}), (\ref{mg:pseudo}) 
to device a perturbative scheme to determine the scaling dimensions $\zeta_{i;r}$ 
of zero modes $\mathcal{Z}_{i;r}$.  Instead of solving perturbatively (\ref{mg:zm}) 
as in \cite{GaKu95,BeGaKu96}, it is possible to read the dependence 
of scaling dimensions upon $\xi$ by looking at how the transition probability density deforms 
statistical conservation laws of the Gaussian theory. 
If statistical conservation laws exist for small but finite 
values of the roughness parameter, the deformation introduces
a logarithmic time dependent prefactor in front of harmonic 
polynomials. Conservation laws at finite $\xi$ must cancel this time 
dependence by containing order by order in $\xi$ suitable logarithmic 
counter-terms in the spatial variables. The constant prefactor of these
terms yields the coefficients of the Taylor expansion of the 
$\zeta_{i;r}$'s. In formulas, the claim is that choosing the state numbers 
$\boldsymbol{J}=(j,\boldsymbol{l})$ so that the action of the transition probability is
within logarithmic accuracy diagonal then
\begin{eqnarray}
\label{mg:scaling}
\tilde{\mathcal{Z}}_{\boldsymbol{J}}=
e^{-t\,\mathcal{M}_{n}}\circ\mathcal{H}_{\boldsymbol{J}}\overset{t \uparrow \infty}{\sim} 
\left(\frac{X}{(\varkappa\,t)^{\frac{1}{2-\xi}}}\right)^{\zeta_{\boldsymbol{J}}(\xi)
-\zeta_{\boldsymbol{J}}(0)}
\mathcal{H}_{\boldsymbol{J}}+\mbox{subleading or amplitude terms}
\end{eqnarray}
Since the perturbative expansion of a statistical conservation law
 $\mathcal{Z}_{i}$ cannot contain any time $t$ dependence, the scaling dimensions of $\mathcal{Z}$ 
and $\tilde{\mathcal{Z}}$ must satisfy
\begin{eqnarray}
\label{}
\mathsf{d}_{\mathcal{Z}}=\mathsf{d}_{\tilde{\mathcal{Z}}}+[\zeta_{\boldsymbol{J}}(\xi)
-\zeta_{\boldsymbol{J}}(0)]
\end{eqnarray}

\section{Calculation of scaling dimensions }
\label{sec:csd}

As solution of (\ref{Hopf:tracer}) at leading order in perturbation theory in $\xi$ 
the transition probability density $\mathcal{P}_{n}$ is
\begin{eqnarray}
\label{csd:tp}
\lefteqn{
\mathcal{P}_{n}=e^{-t\,\mathcal{M}_{n}}=
}
\nonumber\\&&
e^{\frac{\varkappa\,t}{2}\Delta_{n}}-\xi \frac{\varkappa\,t\,\ln m}{2}\Delta_{n}
e^{\frac{\varkappa\,t}{2}\Delta_{n}}+\frac{\xi}{2}\sum_{i\neq j}
(R_{2}^{(1)})_{ij}(e^{\frac{\varkappa\,t}{2}\Delta_{n-2}})_{n/(ij)}+O(\xi^{2})
\end{eqnarray}
The subscript $n/(ij)$ betokens dependence upon all $n$ particles
coordinates but the $i$-th and the $j$-th particle.
\begin{eqnarray}
\label{}
R_{2}^{(1)}(t):=H_{o}(t)\left. \frac{d }{d \xi}\right|_{\xi=0} e^{-t\,\mathcal{M}_{2}}
\end{eqnarray}
is the first order correction to the two-points response function.

\subsection{Diagrammatic expression of $R_{2}^{(1)}$}

$R_{2}^{(1)}$ can be computed by standard diagrammatic techniques of 
statistical field theory (see e.g. \cite{KuMG07}). Denoting for $t\,>\,s$ 
\begin{eqnarray}
e^{\frac{\varkappa\,\Delta_{1}}{2}}=
\setlength{\unitlength}{0.1cm}
\begin{picture}{(0,0)}
\put(0.0,0.0){
\parbox{1.5cm}{\includegraphics[width=1.2cm]
{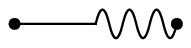}}
}
\put(-1.5,2.0){$\scriptscriptstyle{(\boldsymbol{x},t)}$} 
\put(12.0,2.0){$\scriptscriptstyle{(\boldsymbol{y},s)}$} 
\end{picture}
\hspace{3.0cm}\&\hspace{1.0cm}
D_{(1)}^{\alpha\,\beta}:=\left.\frac{d }{d\xi}\right|_{\xi=0}
D^{\alpha\,\beta}=
\begin{picture}{(0,0)}
\put(0.0,0.0){
\parbox{1.5cm}{\includegraphics[width=1.2cm]
{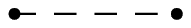}}
}
\end{picture}
\nonumber
\end{eqnarray}
and derivatives by a line perpendicular to response lines, $R_{2}^{(1)}$
is given by
\begin{eqnarray}
\lefteqn{R_{2}^{(1)}(\boldsymbol{x}_{i},\boldsymbol{x}_{j},t|\boldsymbol{y}_{i},\boldsymbol{y}_{j},s)=
\hspace{1.0cm}
\begin{picture}(0,14) 
\put(0.0,0.0){
\parbox{1.7cm}{\includegraphics[width=1.7cm]{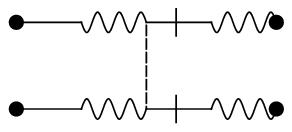}}
}
\put(-20.0,13.0){$\scriptscriptstyle{(\boldsymbol{x}_{i},t)}$} 
\put(-20.0,-10.0){$\scriptscriptstyle{(\boldsymbol{x}_{j},t)}$} 
\put(53.0,13.0){$\scriptscriptstyle{(\boldsymbol{y}_{i},s)}$} 
\put(53.0,-10.0){$\scriptscriptstyle{(\boldsymbol{y}_{j},s)}$} 
\end{picture}}
\nonumber\\&&
\vspace{1.0cm}
\nonumber\\&&
=\int \prod_{r=0}^{2}\frac{d^{d}p_{r}}{(2\,\pi)^{d}}\,
e^{\imath \sum_{l=0}^{2}\boldsymbol{p}_{l}\cdot \boldsymbol{r}_{l}}
e^{-\frac{(t-s)\,\varkappa}{2}\,(\sum_{l=1}^{2}p_l^{2})}\tilde{R}_{2}^{(1)}
(\boldsymbol{p}_{0},\boldsymbol{p}_{1},\boldsymbol{p}_{2},t-s)
\label{diagram:real}
\end{eqnarray}
with
\begin{eqnarray}
\label{diagram:targ}
\tilde{R}_{2}^{(1)}(\boldsymbol{p}_{0},\boldsymbol{p}_{1},\boldsymbol{p}_{2},t)
:=\frac{p_{1\,\alpha}\,p_{2\,\beta}\,
\left(1-e^{-t\,\varkappa\,[p_0^2+\boldsymbol{p}_{0}\cdot(\boldsymbol{p}_{1}-\boldsymbol{p}_{2})]}\right)
}{\varkappa\,\left[p_0^2+\boldsymbol{p}_{0}\cdot(\boldsymbol{p}_{1}-\boldsymbol{p}_{2})\right]}
D^{\alpha\,\beta}_{(1)}(\boldsymbol{p}_{0})
\end{eqnarray}
and
\begin{eqnarray}
\boldsymbol{r}_0:=\boldsymbol{x}_{i}-\boldsymbol{x}_{j}\,, \qquad  
\boldsymbol{r}_1:=\boldsymbol{x}_{i}-\boldsymbol{y}_{i}\,,\qquad  
\boldsymbol{r}_2:=\boldsymbol{x}_{j}-\boldsymbol{y}_{j}
\end{eqnarray}
A (gradient) expansion in the momenta $\boldsymbol{p}_{i}$, 
$\boldsymbol{p}_{j}$ recasts the integral  (\ref{diagram:real}) into the form
\begin{eqnarray}
R_{2}^{(1)}(\boldsymbol{x}_{i},\boldsymbol{x}_{j},t-s|\boldsymbol{y}_{i},\boldsymbol{y}_{j},0)
=\sum_{k=0}^{\infty}\mathcal{V}_{(tr.)}^{(k)}(\boldsymbol{x}_{i},\boldsymbol{x}_{j},t-s)
\,e^{\frac{\varkappa\,(t-s)\,\Delta_{2}}{2}}
(\boldsymbol{x}_{i},\boldsymbol{x}_{j}|\boldsymbol{y}_{i},\boldsymbol{y}_{j})
\label{csd:gradient}
\end{eqnarray}
The vertices $\mathcal{V}_{(tr.)}^{(k)}$'s are homogeneous \emph{normal ordered} differential 
operators of degree $k$. Scaling analysis requires 
to focus on the terms $[\mathcal{V}_{(tr.)}^{(k)}]$ in the $\mathcal{V}_{(tr.)}^{(k)}$'s 
proportional to $\ln (\varkappa\,t)^{-1}$. 
Only the explicit expression of the first few of
them is needed. The reason is the following. Upon inserting 
(\ref{csd:gradient}) into (\ref{csd:tp}), the transition probability 
reduces to
\begin{eqnarray}
\label{diagram:foliation}
e^{-t\,\mathcal{M}_{n}}=e^{\frac{\varkappa\,t}{2}\Delta_{n}}-\xi\left\{
\frac{\varkappa\,t\,\ln m}{2}\Delta_{n}-
\sum_{i\neq j}\sum_{k}\frac{
\mathcal{V}_{(tr.)}^{(k)}(\boldsymbol{x}_{i},\boldsymbol{x}_{j},t)}{2}
\right\}\,e^{\frac{\varkappa\,t}{2}\Delta_{n}}+O(\xi^{2})
\end{eqnarray}
Harmonic polynomials $\mathcal{H}_{\jmath\,\boldsymbol{l}}$ behave as eigenvectors of
unit eigenvalue of the propagator:
\begin{eqnarray}
e^{-t\,\mathcal{M}_{n}}\mathcal{H}_{\jmath\,\boldsymbol{l}}=\mathcal{H}_{\jmath\,\boldsymbol{l}}
+\frac{\xi}{2}\sum_{i\neq j}\sum_{k}\mathcal{V}_{(tr.)}^{(k)}(\boldsymbol{x}_{i},\boldsymbol{x}_{j},t)
\,\mathcal{H}_{\jmath\,\boldsymbol{l}}+O(\xi^{2})
\nonumber
\end{eqnarray}
Suppose that $\mathcal{H}_{\jmath\boldsymbol{l}}$' is particle-permutation and translational invariant, as
it is expected for inertial range tracer correlations.
Suppose furthermore that $\mathcal{H}_{\jmath\boldsymbol{l}}$ contains powers of (components of) 
the variable $\boldsymbol{x}_{i}$ not lower than $k^{\prime}\leq \jmath$ i.e that it has \emph{minimal
homogeneity degree} $k^{\prime}$. Then such polynomial can only be reproduced 
by operators $[\mathcal{V}_{(tr.)}^{(k)}]$ with $k\leq k^{\prime}$. 
The conclusion is that scaling analysis at order $\xi$ requires 
in such a case the spectrum of the homogeneous differential
operator   
\begin{eqnarray}
\label{diagram:ts}
\mathcal{O}_{(tr.)}^{(k^{\prime})}=\frac{1}{2}\sum_{k=0}^{k^{\prime}}
\sum_{i\neq j}[\mathcal{V}_{(tr.)}^{(k)}](\boldsymbol{x}_{i},\boldsymbol{x}_{j})
\end{eqnarray}
Physically, special interest have conserved quantities dominating the
inertial range scaling of structure functions of the tracer. 
Symmetry and analyticity then restrict the corresponding
scaling analysis to the diagonalization of all the 
$\mathcal{O}_{(tr.)}^{(k)}$'s for  $k\leq 2$. At this stage it is worth 
observing that the argument above also applies to the scaling 
analysis of the density field. The action of the transition 
probability on conservation laws of the dual space is
\begin{eqnarray}
\label{}
e^{-t\,\mathcal{M}_{n}^{\dagger}}\mathcal{H}_{\boldsymbol{J}}=\mathcal{H}_{\boldsymbol{J}}+\frac{\xi}{2}
\sum_{i\neq j}\sum_{k}\mathcal{V}_{(de.)}^{(k)}(\boldsymbol{x}_{i},\boldsymbol{x}_{j},t)
\,\mathcal{H}_{\boldsymbol{J}}+O(\xi^{2})
\end{eqnarray}
where the operators $\mathcal{V}_{(de.)}^{(k)}$'s are obtained by
replacing in (\ref{diagram:targ}) $R_{2}^{(1)}$ with
\begin{eqnarray}
\label{}
\tilde{R}_{2}^{(1)\dagger}(\boldsymbol{p}_{0},\boldsymbol{p}_{1},\boldsymbol{p}_{2},t)
:=\frac{(p_{1\,\alpha}+p_{0\,\alpha})(p_{2\,\beta}-p_{0\,\beta})\,
\left(1-e^{-t\,\varkappa\,[p_0^2+\boldsymbol{p}_{0}\cdot(\boldsymbol{p}_{1}-\boldsymbol{p}_{2})]}\right)
}{\varkappa\,\left[p_0^2+\boldsymbol{p}_{0}\cdot(\boldsymbol{p}_{1}-\boldsymbol{p}_{2})\right]}
D^{\alpha\,\beta}_{(1)}(\boldsymbol{p}_{0})
\end{eqnarray}

\subsection{Scaling analysis for the tracer}
\label{sec:st}

The only vertex contributing to (\ref{diagram:ts}) is
\begin{eqnarray}
\label{st:v20}
\mathcal{V}_{(tr.)}^{(2)}(\boldsymbol{x}_{i},\boldsymbol{x}_{j},t)=
\int \frac{d^{d}q}{(2\,\pi)^{d}}\,e^{\imath \boldsymbol{q}\cdot \boldsymbol{x}_{ij}}\,
\frac{\left(1-e^{-t\,\varkappa\,q^2}\right)\,
D_{(1)}^{\alpha\,\beta}(\boldsymbol{q};m)\,
}{\varkappa\,q^{2}}\partial_{x_{i}^{\alpha}}\partial_{x_{j}^{\beta}}
\end{eqnarray}
By (\ref{mg:scaling}) in order to determine scaling it is sufficient 
to extricate the logarithmic asymptotics of (\ref{st:v20}) for
\begin{eqnarray}
\frac{||\boldsymbol{x}_{ij}||^{2}}{\varkappa\,t}\ll\,1
\label{st:approx}
\end{eqnarray}
The technical details of the calculation of the integral have no conceptual 
relevance and are outlined in \ref{ap:integrals}. After some algebra 
(\ref{ap:algebra}), it turns out that 
translational invariant harmonic polynomials of minimal homogeneity degree
two can specify the limit of vanishing $\xi$ of zero modes of 
(\ref{Hopf:tracer}) if they coincide with the harmonic component of the
homogeneous polynomials diagonalizing the operator
\begin{eqnarray}
\label{st:O2}
\lefteqn{\hspace{-1.0cm}
\mathcal{O}_{(tr.)}^{(2)}=\sum_{i\neq j}\frac{[\mathcal{V}^{(2:0)}](\boldsymbol{x}_{i},\boldsymbol{x}_{j})}{2}
=\frac{d+1-2\,\wp}{2\,(d+2)\,(d-1)}
\left\{\qc{O}{n}-\frac{d+(d-2)\,\wp}{d+1-2\,\wp}\qcu{n-1}{n}\right\}}
\nonumber\\&&
-\frac{1}{2\,d}
\left\{\frac{[d+(d-2)\,\wp](d+1-n)[E_{n}+d\,(n-1)]}{(d+2)\,(d-1)\,(n-1)}+
\wp(E_{n}-d)\right\}\,E_{n}
\end{eqnarray}
where $\qc{O}{n}$ and $\qcu{n-1}{n}$ are respectively the representations of
the quadratic Casimir of $SO(d)$ and $SU(n-1)$ on the space of functions
of $n$-particles in $d$-spatial dimensions (see \ref{ap:Casimir}).
As $\qc{O}{n}$, $\qcu{n-1}{n}$ commute, the spectrum of $\mathcal{O}_{(tr.)}^{(2)}$ diagonalized on the
space of translational and permutation invariant homogeneous polynomials 
of degree $\jmath$ can be classified in terms of the Gel’fand-Zetlin patterns 
(see e.g. \cite{Lo70}). In essence, denoting by 
$\boldsymbol{a} = [a_{1} ,\dots , a_{n-2}]$ $n-2$ non-negative integers 
satisfying 
\begin{eqnarray}
\label{st:GZ}
\sum_{i=1}^{n-2}a_{i}=\jmath\hspace{1.0cm} \&\hspace{1.0cm}a_{i}\geq a_{i^{\prime}}
\hspace{0.5cm} \mbox{for}\hspace{0.2cm}i\geq\,i^{\prime}
\end{eqnarray}
the eigenvalues of $\qcu{n-1}{n}$ read
\begin{eqnarray}
\label{st:sun}
\lambda_{SU(n-1)}(\boldsymbol{a})=\sum_{i=1}^{n-2}a_{i}\,(a_{i}-2\,i)
+\frac{\jmath\,[n\,(n-1)-\jmath]}{n-1}
\end{eqnarray}
whilst the Casimir of $\qc{O}{n}$ is
\begin{eqnarray}
\label{}
\lambda_{SO(d)}(\ell)=\ell\,(\ell+d-2)
\end{eqnarray}
for $\ell$ the total (hyper)-angular momentum.
The conclusion is that permutation and translational invariant 
zero modes dominating the inertial range asymptotics of the $n$-point 
tracer correlation functions scale with exponents
\begin{eqnarray}
\label{st:exponents}
\lefteqn{\zeta_{n;1}^{(s)}([a_{1},\dots,a_{n-2}],\ell)=n}
\nonumber\\
&&
+\frac{\xi\,(d+1-2\,\wp)}{2\,(d+2)\,(d-1)}
\left\{\ell\,(\ell+d-2)-\frac{d+(d-2)\,\wp}{d+1-2\,\wp}
\left[\sum_{i=1}^{n-2}a_{i}(a_{i}+n-2\,i)
-\frac{\jmath^{2}}{n-1}\right]\right\}
\nonumber\\
&&
-\frac{\xi\,\jmath}{2\,d}
\left\{\frac{[d+(d-2)\,\wp](d+1-n)[\jmath+d\,(n-1)]}{(d+2)\,(d-1)\,(n-1)}+
\wp\,(\jmath-d)\right\}+O(\xi^{2})
\end{eqnarray}
Of particular interest is the irreducible $SO(d)$ isotropic zero mode
governing the scaling of the structure function. This latter corresponds
to the pattern of highest symmetry $[n,0,\dots,0]$ and scales with exponent
\begin{eqnarray}
\label{}
\zeta_{n;1}^{(s)}([n,0,\dots,0],0)=n-\xi\frac{n\,[d+n+2\,(n-2)\,\wp]}{2\,(d+2)}
+O(\xi^{2})
\end{eqnarray}
a result first obtained in \cite{AdAn98} using operator product expansion
(see also \cite{GaVe00} for discussion).

\subsection{Scaling analysis for the density}
\label{sec:sd}

The adjoint action of the transition probability density brings
about four interaction vertices. The first does not bring about any
differential operation on harmonic polynomials:
\begin{eqnarray}
\label{sa:v0}
\mathcal{V}_{(de.)}^{(0)}(\boldsymbol{x}_{ij},t):=-
\int \frac{d^{d}q}{(2\,\pi)^{d}}\,e^{\imath \boldsymbol{q}\cdot \boldsymbol{x}_{ij}}\,
\frac{q_{\alpha}q_{\beta}\,\left(1-e^{-t\,\varkappa\,q^2}\right)\,
D^{\alpha\,\beta}(\boldsymbol{q};m)}{\varkappa\,q^{2}}
\end{eqnarray}
The other two, after use of the identities
\begin{eqnarray}
\label{}
D^{\alpha\,\beta}(\boldsymbol{q},m)q_{\alpha}q_{\beta}=
-\wp\,D_{\hspace{0.1cm}\alpha}^{\alpha}(\boldsymbol{q},m)\,q^{2} 
\end{eqnarray}
and
\begin{eqnarray}
\label{}
D^{\alpha\,\beta}(\boldsymbol{q},m)q_{\alpha}p_{\beta}=
-\wp\,D_{\hspace{0.1cm}\alpha}^{\alpha}(\boldsymbol{q},m)\,
\boldsymbol{q}\cdot\boldsymbol{p}
\end{eqnarray}
reduce to
\begin{eqnarray}
\label{sd:v1}
\mathcal{V}_{(de.)}^{(1)}(\boldsymbol{x}_{i},\boldsymbol{x}_{j},t)=
-\,\imath\,t\,\wp\,
\int \frac{d^{d}q}{(2\,\pi)^{d}}\,e^{\imath \boldsymbol{q}\cdot \boldsymbol{x}_{ij}}\,
e^{-t\,\varkappa\,q^2}D_{\hspace{0.1cm}\alpha}^{\alpha}(\boldsymbol{q};m)
\boldsymbol{q}\cdot(\partial_{\boldsymbol{x}_{i}}-\partial_{\boldsymbol{x}_{j}})
\end{eqnarray}
and
\begin{eqnarray}
\label{sd:v21}
\mathcal{V}_{(de.)}^{(2:1)}(\boldsymbol{x}_{i},\boldsymbol{x}_{j},t)
=\frac{\wp\,\varkappa\,t^{2}}{2}
\int\frac{d^{d}q}{(2\,\pi)^{d}}e^{\imath\,\boldsymbol{q}\cdot \boldsymbol{x}_{ij}}
e^{-\,t\,\kappa\,q^{2}}D_{\hspace{0.1cm}\alpha}^{\alpha}(\boldsymbol{q},m)
\left[\boldsymbol{q}\cdot(\partial_{\boldsymbol{x}_{i}}-\partial_{\boldsymbol{x}_{j}})\right]^{2}
\end{eqnarray}
The fourth vertex is $\mathcal{V}_{_{(de.)}}^{(2:0)}=\mathcal{V}_{(tr.)}^{(2:0)}$.
As (\ref{sd:v1}) and (\ref{sd:v21}) are not logarithmic in $\varkappa\,t$
\begin{eqnarray}
\label{}
[\mathcal{V}_{(de.)}^{(1)}](\boldsymbol{x}_{i},\boldsymbol{x}_{j})
=[\mathcal{V}_{(de.)}^{(2:1)}](\boldsymbol{x}_{i},\boldsymbol{x}_{j})=0
\end{eqnarray}
the only new contribution relevant for scaling analysis comes from
$\mathcal{V}_{(de.)}^{(0)}$ and is 
\begin{eqnarray}
\label{}
[\tilde{\mathcal{V}}_{(de.)}^{(0)}](\boldsymbol{x}_{ij})=
-\,\frac{\wp\,d}{2}
\end{eqnarray}
(\ref{st:v20}) indicates the existence of zero modes
based at harmonic polynomials of zero degree in the limit $\xi \downarrow 0$.
For these zero modes the operator
\begin{eqnarray}
\label{}
\mathcal{O}_{(de.)}^{(0)}=\frac{1}{2}\sum_{i\neq j}
[\mathcal{V}]^{(0)}(\boldsymbol{x}_{i},\boldsymbol{x}_{j},t)=
-\,\frac{\wp\,d\,n\,(n-1)}{2}
\end{eqnarray}
is diagonal and predicts the scaling dimension \cite{AdAn98,GaVe00}
\begin{eqnarray}
\label{}
\zeta_{n;2}=1-\,\xi\frac{\wp\,d\,n\,(n-1)}{2}+O(\xi^{2})
\end{eqnarray}
The necessary condition that harmonic polynomials of minimal degree two need to
satisfy in order to specify the limit of vanishing $\xi$ of zero modes is to 
specify the harmonic component of the homogeneous polynomials diagonalizing 
the homogeneous operator
\begin{eqnarray}
\label{}
\lefteqn{\hspace{-0.5cm}\mathcal{O}_{(de.)}^{(2)}=-
\frac{\wp\,d\,n\,(n-1)}{2}+
\frac{d+1-2\,\wp}{2\,(d+2)\,(d-1)}
\left\{\qc{O}{n}-\frac{d+(d-2)\,\wp}{d+1-2\,\wp}\qcu{n-1}{n}\right\}}
\nonumber\\&&
-\frac{1}{2\,d}
\left\{\frac{[d+(d-2)\,\wp](d+1-n)[E_{n}+d\,(n-1)]}{(d+2)\,(d-1)\,(n-1)}+
\wp(E_{n}-d)\right\}\,E_{n}
\end{eqnarray}
Again, Gel'fand-Zetlin patterns satisfying (\ref{st:GZ})
provide a natural way to classify the spectrum of $\mathcal{O}_{(de.)}^{(2)}$.
The corresponding spectrum of scaling dimensions is
\begin{eqnarray}
\lefteqn{\zeta_{n;2}^{(s)}([a_{1},\dots,a_{n-2}],\ell)=n-\xi
\frac{\wp\,d\,n\,(n-1)}{2}+\xi
\frac{(d+1-2\,\wp)\,\ell\,(\ell+d-2)}{2\,(d+2)\,(d-1)}
}
\nonumber\\
&&
-\xi
\frac{d+(d-2)\,\wp}{2\,(d+2)\,(d-1)}
\left[\sum_{i=1}^{n-2}a_{i}(a_{i}+n-2\,i)
-\frac{\jmath^{2}}{\jmath-1}\right]
\nonumber\\
&&
-\frac{\xi\,\jmath}{2\,d}
\left\{\frac{[d+(d-2)\,\wp](d+1-n)}{(d+2)\,(d-1)\,(n-1)}[\jmath+d\,(n-1)]+
\wp(\jmath-d)\right\}+O(\xi^{2})
\label{sd:exponents}
\end{eqnarray}

\section{Discussion}
\label{sec:duality}

The general expressions (\ref{st:exponents}), (\ref{sd:exponents}) 
are fairly complicated. It is convenient to discuss their significance 
in the particular non-trivial case $n=4$. This is the case of zero modes 
contributing to fourth order structure and correlation functions. 
Restricting the focus on isotropic zero modes, the Gelf'and-Zetlin patterns
diagonalizing $\mathcal{O}_{(tr.)}^{(2)}$, $\mathcal{O}_{(de.)}^{(2)}$ select two 
harmonic polynomials of degree four
\begin{eqnarray}
\label{duality:irred}
\lefteqn{
\mathcal{H}_{([4,0],0)}\propto\sum_{\{i,i^{\prime}\}}(\boldsymbol{x}_{i}-\boldsymbol{x}_{i^{\prime}})^{4}
}
\nonumber\\&&
-2\sum_{\{\{i, j\},\{i,j^{\prime}\}\}}(\boldsymbol{x}_{i}-\boldsymbol{x}_{j})^{2}
(\boldsymbol{x}_{i}-\boldsymbol{x}_{j^{\prime}})^{2}
+6\sum_{\{\{i, i^{\prime}\},\{j,j^{\prime}\}\}}(\boldsymbol{x}_{i}-\boldsymbol{x}_{i^{\prime}})^{2}
(\boldsymbol{x}_{j}-\boldsymbol{x}_{j^{\prime}})^{2}
\end{eqnarray}
and 
\begin{eqnarray}
\label{duality:red}
\mathcal{H}_{([2,2],0)}\propto -(2\,d+1) 
\sum_{\{i,i^{\prime}\}}(\boldsymbol{x}_{i}-\boldsymbol{x}_{i^{\prime}})^{4}
+(d + 2)\sum_{\{\{i, j\},\{i,j^{\prime}\}\}}(\boldsymbol{x}_{i}-\boldsymbol{x}_{j})^{2}
(\boldsymbol{x}_{i}-\boldsymbol{x}_{j^{\prime}})^{2}
\end{eqnarray}
where the pairs $\{i, i^{\prime}\}$ and $\{j,j^{\prime}\}$ are assumed different, 
as well as the pairs $\{i,j^{\prime}\}$ and $\{j,i^{\prime}\}$. (\ref{duality:irred}), (\ref{duality:red}) 
are the generalization to $d$-dimensions
of the expressions given in \cite{GaKu95} for $d=3$.
They specify the harmonic component of the homogeneous polynomials 
diagonalizing $\mathcal{O}_{(tr.)}^{(2)}$, $\mathcal{O}_{(de.)}^{(2)}$. They differ from the full eigenstate
by linear combinations of \emph{slow modes} i.e. polynomials of 
the form $\phi_{\jmath^{\prime},\boldsymbol{l},\jmath-\jmath^{\prime}}$'s with $\jmath^{\prime}<\jmath$ which are $\mathbb{L}^{2}$-orthogonal to the harmonic 
part on the hypersphere $\mathbb{S}^{8}$ ($\mathbb{S}^{d_n-1}$ in the general case). In this sense, 
harmonic polynomials are replicated by the $\mathcal{O}_{(tr.)}^{(k)}$,$\mathcal{O}_{(de.)}^{(k)}$.  

A second relevant aspect \cite{FaFo05,CeSe05,MaMG09} on which to 
pay attention is represented by the relations interweaving small 
and large scale scaling exponents. Consider the Green function associated 
to the translation invariant transition probability density $P_{n}$. 
The asymptotic expansion
\begin{eqnarray}
M_{n}^{-1}(\boldsymbol{Y},\boldsymbol{Y}^{\prime})=
\sum_{i=0}^{\infty}\left\{
\begin{array}{ll}
\phi_{i;1}(\boldsymbol{Y})\bar{\psi}_{i;1}(\boldsymbol{Y}^{\prime})& Y\,<\,Y^{\prime}
\\
\bar{\psi}_{i;2}(\boldsymbol{Y}^{\prime})\phi_{i;2}(\boldsymbol{Y})& Y\,>\,Y^{\prime}
\end{array}
\right.,\hspace{0.5cm}
\bar{\psi}_{i;r}(\boldsymbol{Y}):=\int_{0}^{\infty}dt\,\psi_{i;r}(\boldsymbol{Y},t)
\end{eqnarray}
dictates
\begin{eqnarray}
\label{}
\mathsf{d}_{\bar{\psi}_{n;1}}+\mathsf{d}_{\phi_{n;1}}=
\mathsf{d}_{\bar{\psi}_{n;2}}+\mathsf{d}_{\phi_{n;2}}=[2-\xi-d(n-1)]\,\mathsf{d}_{x}
\end{eqnarray}
Bearing in mind the physical interpretation of the adjoint action
of $P_{n}$ these relations state that the knowledge of the scaling
dimensions of inertial range zero modes of the tracer field entails that
of large scale zero modes of the density field and vice-versa.
For example, denoting with superscripts $(s)$ and $(l)$ respectively small
and large scale exponents, the isotropic irreducible and reducible 
zero modes of the tracer four point correlation function give 
\begin{eqnarray}
\label{}
\lefteqn{
\begin{array}{l}
\zeta_{4;1}^{(s)}([4,0],0)=4-2\,\xi \frac{d+4+4\,\wp}{d+2}+O(\xi^{2})
\\
\zeta_{4;1}^{(s)}([2,2],0)=4-2\,\xi \frac{d-2+\wp}{d-1}+O(\xi^{2})
\end{array}
}\nonumber\\&&
\hspace{0.5cm}\Rightarrow\hspace{0.5cm}
\begin{array}{l}
\zeta_{4;2}^{(l)}([4,0],0)=-(3\,d+2)+\xi\frac{d+6+8\,\wp}{d+2}+O(\xi^{2})
\\
\zeta_{4;2}^{(l)}([2,2],0)=-(3\,d+2)+\xi\frac{d-3+2\,\wp}{d-1}+O(\xi^{2})
\end{array}
\end{eqnarray}
whilst the corresponding object for the density yield
\begin{eqnarray}
\label{}
\lefteqn{
\begin{array}{l}
\zeta_{4;2}^{(s)}([4,0],0)=4-2\,\xi \frac{d+4+\wp\,[4+3\,d\,(d+2)]}{d+2}+O(\xi^{2})
\\
\zeta_{4;2}^{(s)}([2,2],0)=4-2\,\xi \frac{d-2+\wp\,[1+3\,d\,(d-1)]}{d-1}+O(\xi^{2})
\end{array}
}
\nonumber\\&&
\hspace{0.5cm}\Rightarrow\hspace{0.5cm}
\begin{array}{l}
\zeta_{4;1}^{(l)}([4,0],0)=-(3\,d+2)+\xi\frac{d+6+2\,\wp\,[4+3\,d(d+2)]}{d+2}
+O(\xi^{2})
\\
\zeta_{4;1}^{(l)}([2,2],0)=-(3\,d+2)+\xi\frac{d-3+2\,\wp\,[1+3\,d\,(d-1)]}{d-1}
+O(\xi^{2})
\end{array}
\end{eqnarray}
As expected, in the incompressible limit of vanishing $\wp$ the relations 
become self-dual. Large scale zero modes may become manifest in the power law 
decay of certain statistical indicators at large point separations if the 
integral scale of the velocity field is much larger than the one of the
forcing \cite{CeSe05,MaMG09}. 
A final observation is that a direct determination of scaling dimensions 
of large scale zero modes is naturally achieved by studying the time dependent 
martingales $\psi_{j\,\boldsymbol{l}\,0}$ rather their stationary 
counterparts $\bar{\psi}_{j\,\boldsymbol{l}\,0}$ which are not martingales.
Within the approximation (\ref{st:approx}), straightforward algebra yields
\begin{eqnarray}
\label{duality:time}
\lefteqn{\int d^{d_{n}}Y^{\prime}\,P_{n}(\boldsymbol{Y},\boldsymbol{Y}^{\prime},t-s) 
\psi_{n\boldsymbol{l},0}(\boldsymbol{Y}^{\prime},s)
\overset{\frac{R^{2}}{\varkappa\,\tau}\ll\,1}{=}\psi_{n\boldsymbol{l},0}(\boldsymbol{Y},t)+}
\nonumber\\&&
\xi\,\psi_{n\boldsymbol{l},0}(\boldsymbol{Y},t)\left\{-
\frac{1}{2}\ln (\sqrt{\varkappa\,\tau})\,\left(\frac{\dn}{2}+n\right)
+\ln \frac{Y}{\sqrt{\varkappa\,\tau }}\frac{1}{\mathcal{H}_{\jmath\boldsymbol{l}}}
\mathcal{O}^{(2)}\mathcal{H}_{\jmath\boldsymbol{l}}
\right\}+\dots+O(\xi^{2})
\end{eqnarray} 
where $\tau=t-s$ and dots stand for slow modes. Scaling prediction for
time-dependent martingales can be then read from the prefactor of the
logarithmic counter-terms needed to compensate the the $\ln(\varkappa\,\tau )^{-1}$
dependence in (\ref{duality:time}).

\section{Conclusions}
\label{sec:end}

The technique illustrated is strongly reminiscent of the operator product 
expansion \cite{Zinn} previously also used to perform systematic 
calculations in the Kraichnan model (see e.g. \cite{AdAnVa98,AdAn98,AdAnBaKaVa01,KuMG07} 
and references therein). The differences are that the role of momentum cut-off's 
is played by the time and that instead of bases of abstract 
operator valued fields (composite operators) it makes use of a 
basis of harmonic polynomials or shapes $\mathbb{L}^{2}$ complete on the $\mathbb{S}^{d_{n}-1}$ hypersphere. 
These objects have a direct geometrical meaning and can be used in numerical
simulations close to controlled or phenomenological Gaussian limits
to test scaling properties as it was done in \cite{MaMG09}.
A final observation is that, at least in principle, the technique 
does not require linearity of the statistical field theory in order 
to be applicable. The reason is the following. In the thermodynamic formalism of field theory \cite{Zinn}, 
connected $n$-point correlation functions $W^{(n)}$ are reconstructed by solving Hopf-like equations of
the form
\begin{eqnarray}
U^{(2)}\,W^{(n)}=\mathcal{F}_{n}(W^{(n-1)},...,W^{(1)})
\label{end:Legendre}
\end{eqnarray}
where $U^{(2)}$ stands for the set of \emph{proper vertices} of order two acting on 
$n$-point connected correlation functions. The functionals $\mathcal{F}_{n}$'s 
in (\ref{end:Legendre}) are specified by functional derivatives of the Legendre 
transform connecting the free energy $W$ (i.e. the
generating function of connected correlations) to the thermodynamic potential (i.e. the
generating function of proper vertices). Thus the $\mathcal{F}_{n}$'s depend upon all 
proper vertices of order less or equal to $n$. Once these latter ones are given, (\ref{end:Legendre}) 
defines a solvable hierarchy for the connected correlations. Thinking of the $U^{(n)}$'s as 
pseudo-differential operators, the generalization of the method presented in this paper 
consists in probing the scaling of "martingales" associated to the generalized 
propagators $U^{(2)-1}$ acting on $n$-particle functions. It must be, however, emphasized
that deriving expressions of proper vertices is in general a very non-trivial task that 
can be most often carried out only when perturbative methods are applicable.  

\section{Acknowledgements}

Discussions with J. Bec and A. Mazzino are gratefully acknowledged. 
This work was supported by the center of excellence “Analysis and Dynamics” 
of the Academy of Finland.

\appendix

\section{Outline of the evaluation of the integrals}
\label{ap:integrals}

A convenient method to remove cut-offs from integrals associated to Feynman-diagram
is to take the Mellin transform with respect to the cut-off (see e.g. \cite{Zinn}).
For the velocity correlation this means 
\begin{eqnarray}
\label{integrals:Mellin}
\tilde{D}_{z}^{\alpha\,\beta}(\boldsymbol{x},m)=-\frac{D_{0}\,\xi\,m^{z-\xi}\,\hat{c}(z;\xi)}{z-\xi}
\int\frac{d^{d}p}{(2\,\pi)^{d}}
e^{\imath\,\boldsymbol{p\cdot x}}\frac{\Pi(\boldsymbol{\hat{p},\wp})}{p^{d+z}}
\end{eqnarray} 
with $\Pi(\boldsymbol{\hat{p},\wp})=(1-\wp)\delta^{\alpha\beta}-(1-d\,\wp)p^{\alpha}p^{\beta}/p^{2}$
and
\begin{eqnarray}
\label{Mellin:cfou}
\frac{\hat{c}(z;\xi)}{z-\xi}:=\int_{0}^{\infty}\frac{dw}{w}\frac{\chi(w^{2})}{w^{z-\xi}}
\hspace{1.0cm} \& \hspace{1.0cm} \hat{c}(0;\xi)=1
\end{eqnarray} 
All non-universal information of the cut-off function is stored in the residues 
of the function $\hat{c}$ for $\Re z\,\geq\,2$. These residues are, however, of no relevance 
for universal properties of the system and further description of them is needed. 
The scope of this appendix is to apply the Mellin transform 
to the evaluation of the logarithmic asymptotics 
of the integrals introduced in the main text. The interested reader is referred to
\cite{KuMG07} for more details on this technique.

\subsection{Evaluation of $\mathcal{V}_{(de.)}^{(0)}$}

After taking the Mellin transform with respect to the two cut-off scales
$m$ and $\varkappa\,t$ (\ref{sa:v0}) becomes
\begin{eqnarray}
\label{}
\lefteqn{
\tilde{\mathcal{V}}_{(de.);z,\zeta}^{(0)}
=
}
\nonumber\\&&
-\frac{D_{o}\,m^{z}\,\wp\,(d-1)\,\hat{c}(z;0)}{\varkappa\,z}
\int\frac{d^{d}q}{(2\,\pi)^{d}}\,e^{\imath\boldsymbol{q \cdot x}}
\left\{\frac{1}{q^{d+z}}-\int\frac{d\zeta}{(2\,\pi\,\imath)}
\frac{\Gamma(-\zeta)\,(\varkappa \tau)^{\zeta}}{q^{d+z-2\,\zeta}}\right\}
\end{eqnarray}
The momentum integral can be performed for
\begin{eqnarray}
\label{}
\Re (z-2\,\zeta)\,<\,0\hspace{0.5cm} \&\hspace{0.5cm} \Re\,z\,<\,0
\end{eqnarray}
The asymptotics for
\begin{eqnarray}
\label{integrals:asympt}
\frac{x^{2}}{\varkappa\,\tau}\,\ll\,1
\end{eqnarray}
corresponds to a shift to the left of the contour of the inverse Mellin 
transform in $\zeta$. In order to extricate the leading order of the
asymptotics it is sufficient to approximate the inverse transform with
the residue of the first pole which is encountered for $\Re \zeta=z/2$ ($\Re z\,<\,0$):
\begin{eqnarray}
\label{}
\tilde{\mathcal{V}}_{z}^{(0:0)}=\frac{\wp\,d\,
\,(m\,x)^{z}}{z}\left\{
\frac{\Gamma\left(\frac{d}{2}\right)\Gamma\left(1-\frac{z}{2}\right)}
{2^{z}\,z\,\Gamma\left(\frac{d+z}{2}\right)}-
\frac{\Gamma\left(1-\frac{z}{2}\right)}{z}
\left(\frac{\varkappa\,\tau}{x^{2}}\right)^{\frac{z}{2}}\right\}+\mbox{sub-leading}
\end{eqnarray}
Finally the limit of vanishing $m$ requires shifting the contour of 
the inverse Mellin transform in $z$ towards positive values of $\Re\,z$. The result is
\begin{eqnarray}
\label{}
\mathcal{V}^{(0:0)}\left(\boldsymbol{x},t\right)
=-\,\frac{\xi\,\wp\,d}{2}\,\ln\frac{x^{2}}{\varkappa\,\tau}+\mbox{sub-leading}
\end{eqnarray}  

\subsection{Evaluation of $\mathcal{V}_{(tr.)}^{(2)}$}

Proceeding as above (\ref{st:v20}) becomes
\begin{eqnarray}
\label{int20:int20}
\lefteqn{\tilde{\mathcal{V}}_{(tr.);z\zeta}^{(2:0)}(\boldsymbol{x}_{12})=
-\frac{D_{o}\,m^{z}\,\hat{c}(z,0)}{z}\times}
\nonumber\\&&
\int \frac{d^{d}q}{(2\,\pi)^{d}}\,e^{\imath \boldsymbol{q}\cdot \boldsymbol{x}_{12}}\,
\left\{1-\int\frac{d\zeta}{(2\,\pi\,\imath)}
\frac{\Gamma(-\zeta)\,(\kappa\,\tau)^{\zeta}}{q^{-2\,\zeta}}\right\}
\frac{\Pi^{\alpha\,\beta}(\boldsymbol{\hat{q}},\wp)}{q^{d+z+2}}
\partial_{x_{1}^{\alpha}}\partial_{x_{2}^{\beta}}
\end{eqnarray}
The momentum integral exists for 
\begin{eqnarray}
\label{}
\Re \zeta= z+2\,<\,0\hspace{1.0cm}\&\hspace{1.0cm}\Re z\,<\,-2
\end{eqnarray}
The asymptotic expression of the logarithmic part of the vertex in 
the limit (\ref{integrals:asympt}) is obtained by approximating 
the inverse Mellin transform in $\zeta$ with the residues of the
poles for $\Re \zeta=(z+2)/2$ and $\Re \zeta=z/2$. The resulting pole
in $z$ is double with residue 
\begin{eqnarray}
\label{integrals:v20}
\lefteqn{
\mathcal{V}_{(tr.)}^{(2)}(\boldsymbol{x}_{i},\boldsymbol{x}_{j},t)
=-\varkappa\,t \ln\sqrt{m^{2}\varkappa\,t}
\,\partial_{\boldsymbol{x}_{i}}\cdot\partial_{\boldsymbol{x}_{j}}
}
\nonumber\\&&
+\frac{2\,(1-\wp)+d-1}
{4\,(d+2)\,(d-1)}\ivec^{\alpha\,\beta}(\boldsymbol{\hat{x}}_{ij};2,\wp)
\ln\frac{x_{ij}^{2}}{\varkappa\,t}\,x_{ij}^{2}\partial_{x_{i}^{\alpha}}\partial_{x_{j}^{\beta}}
+\mbox{sub-leading}
\end{eqnarray}
with
\begin{eqnarray}
\ivec^{\alpha\,\beta}(\boldsymbol{x};z,\wp):=\delta^{\alpha\,\beta}-
\frac{z(1-d\,\wp)}{d+z(1-\wp)-1} \frac{x^{\alpha} x^{\beta}}{x^{2}}
\label{st:projector}
\end{eqnarray}
Note that whenever acting on translational invariant functions
\begin{eqnarray}
\label{}
\lefteqn{
\sum_{i\neq j}\mathcal{V}_{(tr.)}^{(2)}(\boldsymbol{x}_{i},\boldsymbol{x}_{j},t)=
\varkappa\,t \ln\sqrt{m^{2}\varkappa\,t}\,\Delta_{n}
}
\nonumber\\
\nonumber\\&&
+\frac{2\,(1-\wp)+d-1}
{4\,(d+2)\,(d-1)}\sum_{i\neq j}\ivec^{\alpha\,\beta}(\boldsymbol{\hat{x}}_{ij};2,\wp)
\ln\frac{x_{ij}^{2}}{\varkappa\,t}\,x_{ij}^{2}\partial_{x_{i}^{\alpha}}\partial_{x_{j}^{\beta}}
+\mbox{sub-leading}
\end{eqnarray}

\section{Definition of quadratic Casimir}
\label{ap:Casimir}

Let $i$ be the particle label $i=1,\dots,n$ and $\alpha,\beta$
be index of vector components of individual particles in 
$\mathbb{R}^{d}$. Upon defining
\begin{eqnarray}
\label{}
\mathfrak{g}_{i}^{\alpha\,\beta}:=x_{i}^{\alpha}\partial_{x_{i}^{\beta}}
\hspace{0.5cm}\& \hspace{0.5cm}
\mathfrak{e}_{i}:=\mathfrak{g}_{i\hspace{0.1cm}\alpha}^{\alpha}
\end{eqnarray}
the dilation operator $E_{n}$ acting on the n-particle space 
$\mathbb{R}^{n\,d}$ is
\begin{eqnarray}
\label{Casimir:dilation}
E_{n}=\sum_{i=1}^{n}\mathfrak{e}_{i}
\end{eqnarray}
On the same space, the representation of the quadratic Casimir of
 $SU(d)$ is 
\begin{eqnarray}
\label{}
\qc{U}{n}=\sum_{i\,j}\left(\mathfrak{g}_{i}^{\alpha\,\beta}-\frac{\delta^{\alpha\,\beta}}{d}
\mathfrak{e}_{i}\right)
\left(\mathfrak{g}_{j;\beta\,\alpha}-\frac{\delta_{\beta\,\alpha}}{d}\mathfrak{e}_{j}\right)
=\sum_{i\,j}\tr \mathfrak{g}_{i}\mathfrak{g}_{j}-\frac{E_{n}^{2}}{d}
\end{eqnarray}
Using the dilation operator, this latter can be also written in 
terms of the quadratic Casimir of $SU(n-1)$:
\begin{eqnarray}
\label{Casimir:sun}
\qc{U}{n}=\qcu{n-1}{n}+\frac{d+1-n}{d\,(n-1)}E_{n}\,[E_{n}+d(n-1)]
\end{eqnarray}
The generators of $SO(d)$ for the $i$-th particle in $d$ dimensions are
\begin{eqnarray}
\label{}
\mathfrak{l}_{i}^{\alpha\,\beta}:=\mathfrak{g}_{i}^{\alpha\,\beta}-\mathfrak{g}_{i}^{\alpha\,\beta}
\end{eqnarray}
The generators of $SO(d)$ for $n$ particles are additive functions 
of the generators of $SO(d)$ for single particle:
\begin{eqnarray}
L^{\alpha\,\beta}_{n}=\sum_{i=1}^{N} \mathfrak{l}_{i}^{\alpha\,\beta}
\end{eqnarray}
The quadratic Casimir is
\begin{eqnarray}
\qc{O}{n}:=\frac{1}{2}L^{\alpha\beta}L_{\alpha\beta}=
\sum_{i j=1}^{n}\left\{\tr \left(\ud{i}\ud{j}\right)\right\}-\tr \left(\ud{i}\ud{j}^{t}\right)
\end{eqnarray}
A excellent introduction to group theoretic methods for many-body 
systems can be found in \cite{Lo70}.

\section{Derivation of (\ref{st:O2})}
\label{ap:algebra}

Omitting the Laplacian term as it vanishes on harmonic 
polynomials $[\mathcal{V}_{(tr.)}^{(2)}]$ reduces to
\begin{eqnarray}
\label{algebra:O2}
\lefteqn{\sum_{i\neq j}[\mathcal{V}_{(tr.)}^{(2)}](\boldsymbol{x}_{i},\boldsymbol{x}_{j})=}
\nonumber\\&&
\frac{d+2(1-\wp)-1}{2\,(d+2)\,(d-1)}\sum_{ij}\left[x_{ij}^{2}
\partial_{\boldsymbol{x}_{i}}\cdot\partial_{\boldsymbol{x}_{j}}-
\frac{2\,(1-d\,\wp)}{d+2(1-\wp)-1}x_{ij}^{\alpha}x_{ij}^{\beta}\partial_{x_{i}^{\alpha}}
\partial_{x_{j}^{\beta}}\right]
\end{eqnarray}
Unfolding the differences in $\boldsymbol{x}_{ij}=\boldsymbol{x}_{i}-\boldsymbol{x}_{j}$ and dropping differential
terms vanishing on translation invariant functions, yields
\begin{eqnarray}
\label{}
\lefteqn{\sum_{ij}\left[x_{ij}^{2}
\partial_{\boldsymbol{x}_{i}}\cdot\partial_{\boldsymbol{x}_{j}}-
\frac{2\,(1-d\,\wp)}{d+2\,(1-\wp)-1}x_{ij}^{\alpha}x_{ij}^{\beta}\partial_{x_{i}^{\alpha}}
\partial_{x_{j}^{\beta}}\right]}
\nonumber\\
&&
=-\,2\,\sum_{ij}\left\{\boldsymbol{x}_{i}\cdot\boldsymbol{x}_{j}
\partial_{\boldsymbol{x}_{i}}\cdot\partial_{\boldsymbol{x}_{j}}-
\frac{2\,(1-d\,\wp)}{d+2\,(1-\wp)-1} \frac{x_{i}^{\alpha}x_{j}^{\beta}+
x_{j}^{\alpha}x_{i}^{\beta}}{2}\partial_{x_{i}^{\alpha}}
\partial_{x_{j}^{\beta}}\right\}
\end{eqnarray}
Applying the definitions of appendix~\ref{ap:Casimir}, finally 
gives
\begin{eqnarray}
\label{}
\lefteqn{\sum_{i\neq j}\boldsymbol{x}_{ij}\cdot\partial_{\boldsymbol{x}_{ij}}
\mathcal{V}^{(2:0)}(\boldsymbol{x}_{ij})
=\frac{(d+1-2\,\wp)}{(d+2)\,(d-1)}}
\nonumber\\&&
\left\{\qc{O}{n}-\frac{d+(d-2)\,\wp}{d+1-2\,\wp}\qc{U}{n}
-\frac{(d-1)(d+2)\wp}{d\,(d+1-2\,\wp)}E_{n}(E_{n}-d\,n)\right\}
\end{eqnarray}
The equality (\ref{Casimir:sun}) permits to express this result
in terms of $\qcu{n-1}{n}$.


\end{document}